%% file: BPH-21-003_temp.tex
\pdfoutput=1
\documentclass[11pt,twoside,a4paper,cmspaper,final,collab]{cms-tdr}
\def\svnVersion{b252d54}\def\svnDate{2024/03/18}\def\cmsCernNoTag{CERN-EP-2023-109}\def\cmsCernDate{\today}\def\cmsMessage{Published in Physical Review Letters as \href{http://dx.doi.org/10.1103/PhysRevLett.132.111901}{\doi{10.1103/PhysRevLett.132.111901}.}}
\begin{document}\cmsNoteHeader{BPH-21-003}

\renewcommand{\thetable}{\Roman{table}}
\captionsetup[table]{name=TABLE.}
\captionsetup[figure]{name=FIG.}
\captionsetup[table]{labelsep=period}
\captionsetup[figure]{labelsep=period}

\newcommand{\HELACONIA} {{\textsc{helac-o}{\footnotesize nia}}\xspace}
\newcommand{\CASCADE}   {{\textsc{cascade}}\xspace}
\newcommand{\JHUGEN}    {{\textsc{jhug}{\footnotesize en}}\xspace}
\newcommand{\myGEANT} {{\textsc{geant\footnotesize{4}}}\xspace}
\newcommand{\myLone}   {{\textit{L}1}\xspace}

\newcommand{\JHU}       {{\textsc{JHU}}\xspace}
\newcommand{\JJ}        {\ensuremath{\PJGy\PJGy}\xspace}
\newcommand{\Xss}       {\HepParticleResonance{X}{6600}{}{}\Xspace}
\newcommand{\Xsn}       {\HepParticleResonance{X}{6900}{}{}\Xspace}
\newcommand{\Xst}       {\HepParticleResonance{X}{7100}{}{}\Xspace}
\newcommand{\Xtest}     {\HepParticleResonance{X}{3872}{}{}\Xspace}
\newlength\cmsTabSkip\setlength{\cmsTabSkip}{1ex}
   
\ifthenelse{\boolean{cms@external}}
{\providecommand{\suppMaterial}{the Supplemental Material~\cite{suppMatBib}}}
{\providecommand{\suppMaterial}{Appendix~\ref{app:suppMat}\nocite{suppMatBib}}}

\ifthenelse{\boolean{cms@external}}
{\providecommand{\suppMaterialB}{the Supplemental Material~\cite{suppMatBib}}}
{\providecommand{\suppMaterialB}{Appendix~\ref{app:suppMatB}\nocite{suppMatBib}}}

\ifthenelse{\boolean{cms@external}}{\providecommand{\lhcbfitreference}{in }}{\providecommand{\lhcbfitreference}{in Fig. A.1 of }}
\ifthenelse{\boolean{cms@external}}{\providecommand{\NA}{\ensuremath{\cdots}\xspace}}{\providecommand{\NA}{\ensuremath{\text{---}}\xspace}}

\cmsNoteHeader{BPH-21-003}
\title{New Structures in the \texorpdfstring{\JJ}{J/psi J/psi} Mass Spectrum in Proton-Proton Collisions at \texorpdfstring{$\sqrt{s} = 13\TeV$}{sqrt(s) = 13 TeV}}

\date{\today}
 
\abstract{
A search is reported for near-threshold structures in the \JJ invariant mass spectrum produced in proton-proton 
collisions at $\sqrt{s} = 13\TeV$ from data collected by the CMS experiment, corresponding to an 
integrated luminosity of $135\fbinv$. 
Three structures are found, and a model with quantum interference among these structures provides a good 
description of the  data.
A new structure is observed with a local significance above 5 standard deviations at a mass of
$6638^{+43}_{-38} \stat ^{+16}_{-31} \syst\MeV$. 
Another structure with even higher significance is found at a mass of
$6847^{+44}_{-28} \stat ^{+48}_{-20} \syst\MeV$, which is consistent with the \Xsn resonance 
reported by the LHCb experiment and confirmed by the ATLAS experiment.
Evidence for another new structure, with a local  significance of 4.7 
standard deviations, is found at a mass of $7134^{+48}_{-25} \stat ^{+41}_{-15} \syst\MeV$.
Results are also reported for a model without interference,  
which does not fit the data as well and shows mass shifts up to 150\MeV relative to the model 
with interference.
}

\hypersetup{
pdfauthor={CMS Collaboration},
pdftitle={Observation of new structure in the J/psi J/psi mass spectrum in proton-proton collisions at sqrt(s) = 13 TeV},
pdfsubject={CMS},
pdfkeywords={CMS, exotic meson, exotic hadron, tetraquark}}
     
\maketitle

The prospect of exotic hadrons---states other than $\PQq\PAQq$ or 
$\PQq\PQq\PQq$---dates back to Gell-Mann's 
seminal 1964 paper~\cite{Gell-Mann:1964ewy}.
However, interest in exotics, both theoretical and experimental, wavered for decades.
The discovery of the charmoniumlike \Xtest state
by the Belle Collaboration in 2003~\cite{Belle:2003nnu}
propelled exotics from speculative chimeras 
to the frontier of hadronic physics~\cite{ali_maiani_polosa_2019}.
While many tetraquark candidates containing heavy quarks (\PQc, \PQb) are now known, 
questions still abound, including 
which of
these are truly exotic hadrons~\cite{ali_maiani_polosa_2019} 
and, if they are bound states, what is their internal structure 
(\eg, molecules, bound states of diquarks, \etc~\cite{Brambilla:2019esw,Nielsen:2018ytt,Nielsen:2018uyn}).

Beginning with the \PJGy discovery in 1974~\cite{E598:1974sol, SLAC-SP-017:1974ind}, 
heavy quarkonia 
brought clarity to the quark model. 
Similarly, exotic states composed entirely of heavy quarks 
may offer analogous insights.
Exotic states decaying into the \JJ and $\PGU\PGU$ channels are particularly
promising experimentally 
because of the efficient triggering and reconstruction of muonic channels.
The possibility of structure appearing in these channels is enhanced by an empirical
pattern first observed with light mesons:
the mass spectra of two vector states with the same isospin (\eg, $\PGf\PGf$ and $\PGo\PGf$)
have near-threshold enhancements, whereas systems of two vector mesons of different isospin
(\eg, $\PGf\PGr$ and $\PGo\PGr$) have no structure~\cite{Yi:2013yuf}.
Such a behavior has been corroborated by the enhancements observed
in $\PJGy\PGo$~\cite{BaBar:2010wfc} and $\PJGy\PGf$~\cite{LHCb:2021uow} spectra.

The first searches for all heavy quark states were in the all-bottom sector and 
were inconclusive~\cite{LHCb:2018uwm,CMS:2020qwa}.
In 2020, the LHCb Collaboration reported structure 
in the \JJ channel, which consisted of a broad enhancement just above the \JJ 
mass and a narrow peak designated as \Xsn~\cite{LHCb:2020bwg}.
The nature of the broad structure was uncertain, but the narrow peak was 
interpreted to be an all-charm tetraquark 
state~\cite{Wang:2020wrp,Chen:2020xwe,Bedolla:2019zwg,Jin:2020jfc,liu:2020eha,Liu:2021rtn,Zhu:2020xni,Giron:2020wpx}.
Considerations of such states
date back to 1976~\cite{Iwasaki:1976cn,Chao:1980dv}
and have recently become the object of renewed interest~\cite{Berezhnoy:2011xn,Wu:2016vtq,Bai:2016int,Wang:2017jtz,Anwar:2017toa,Richard:2017vry,Esposito:2018cwh,Karliner:2016zzc,Liu:2019zuc,Chen:2016jxd,Wang:2019rdo}.
However, nontetraquark interpretations of this new structure
have also been proposed~\cite{Gong:2020bmg,Wan:2020fsk,Wang:2020tpt}.

Here we report on 
the \JJ invariant mass spectrum from proton-proton ($\Pp\Pp$) collisions 
as recorded with the CMS detector,
where the \PJGy mesons are reconstructed from \MM pairs.
The data sample corresponds to an integrated luminosity of
135\fbinv~\cite{CMS:2021xjt,CMS:2018elu,CMS:2019jhq} 
at a center-of-mass energy of 13\TeV. 

The central feature of the CMS apparatus is a superconducting solenoid 
of 6\unit{m} internal diameter, providing a magnetic field of 3.8\unit{T}. 
Within the solenoid volume are a silicon pixel and strip tracker, 
a lead tungstate crystal electromagnetic calorimeter, 
and a brass and scintillator hadron calorimeter, 
each composed of a barrel and two end cap sections. 
Muons are detected in gas-ionization chambers
embedded in the steel flux-return yoke outside the solenoid.
A more detailed description of the CMS detector
can be found in Ref.~\cite{CMS:2008xjf}.

Several Monte Carlo (MC) samples are used to model 
the detector mass resolution and efficiency of signals.
As simulations of tetraquarks states are not available and the analysis is
independent of signal kinematic details, we use the production of other
particles to simulate the signal, with the masses defined appropriately.
We use $J^P = 0^+$ mesons for a variety of masses 
decaying to \JJ as the signal 
and simulate their production via gluon-gluon interactions.
We model signal particles by \PGcbDzP{1P} production
using the \PYTHIA event generator (version 8.230)~\cite{Sjostrand:2014zea}. 
As an alternate production model, signal is also simulated by generating
a sample with "Higgs bosons" decaying to $\PZ\PZ^{(*)}$,
where the $\PZ^{(*)}$ is redefined as a \PJGy, using 
the \JHUGEN event generator (version 7.40)~\cite{1001.3396,Gritsan_2020}.
To understand the detector mass resolution, 
the samples are generated with a negligible natural width.

Major sources of background, \ie, selected events that are not from a resonant decay to 
two \PJGy candidates, are expected to originate from 
two genuine \PJGy decays arising either from a single 
parton-parton collision, \ie, nonresonant single-parton scattering 
(NRSPS)~\cite{Qiao_2010,Lansberg_2013,Lansberg_2015,Sun_2016,Baranov_2016,Likhoded_2016}, or a pair of 
parton-parton interactions in a single $\Pp\Pp$ collision, \ie, double-parton scattering 
(DPS)~\cite{Calucci_1999,Del_Fabbro_2001,Del_Fabbro_2001,Kom_2011,Baranov_2013,d_Enterria_2013,He_2015,Borschensky_2017}. 
The NRSPS contribution is expected to dominate the DPS contribution near the 
\JJ threshold, with the DPS contribution dominating at masses above 11 \GeV. 
To simulate these backgrounds we use 
the \PYTHIA generator.
The next-to-leading order \CASCADE generator~\cite{Jung_2010} 
and the next-to-next-to-leading order \HELACONIA generator~\cite{Shao_2013,Lansberg_2015,Lansberg_2013,shao2015helaconia} 
are also used to simulate alternative NRSPS shapes
for the estimation of the systematic 
uncertainty in the background.
Direct production, as well as feed-down
processes where \PJGy mesons arise as decay products, are included.
Feed-down processes from double-charmonium
resonances~\cite{Wang:2020tpt} are generated 
using the \PYTHIA generator separately.
Generated events are processed through the full CMS detector simulation,
which utilizes the \myGEANT toolkit~\cite{GEANT4:2002zbu}.

Events of interest are selected using a two-tiered trigger system~\cite{CMS:2016ngn}. 
The first level \myLone, composed of custom hardware processors, uses information 
from the calorimeters and muon detectors to select events. The second level, known as the 
high-level trigger (HLT), consists of a farm of processors running a faster version of 
the full event reconstruction software, which reduces the event rate to around~1\unit{kHz}.
The data collected in 2016 correspond to an integrated luminosity of 36.3\fbinv, and were obtained with 
an \myLone trigger that required at least three muon candidates 
to balance the requirements of high efficiency for the signal and a low trigger rate.
The HLT required that the pseudorapidity of 
each muon satisfies $\abs{\eta}<2.5$ 
and that a muon pair candidate is composed of oppositely charged muon candidates with an invariant mass
between 2.95 and 3.25\GeV, 
a distance of closest approach between the two muons less than 0.5\cm,
and a fit to a common vertex with a $\chi^2$ probability greater than 0.5\%.
The data collected in 2017--2018 correspond to an integrated luminosity of 98.6\fbinv and utilized 
an \myLone trigger requiring at least three muons with at least two muons having 
$\pt > 3$\GeV and one  of the two muons having $\pt > 5$\GeV,  
and at least one oppositely charged pair of muons with invariant mass below 9\GeV.
The HLT criteria replicated those of 2016
with the additional 
requirement
that the two muons with invariant mass between 2.95 and 3.25\GeV 
each have $\pt>3.5\GeV$.

The off-line selection criteria were finalized before unblinding the signal
region (defined as invariant mass below 7.8\GeV) and are based in part on an 
examination of data collected in 2011 and 2012.
It requires at least four muons, each of them having $\pt > 2\GeV$, 
$\abs{\eta} < 2.4$, and satisfying the 
``soft muon identification'' requirements~\cite{CMS:2012nsv}. 
Pairs of oppositely charged muons are used to form \PJGy candidates, and these are 
required to have $\pt>3.5\GeV$. 
To confirm the HLT requirement, each muon pair is fit to a vertex, requiring a fit
probability greater than 0.5\%. 
Combinations of four muons from two \PJGy candidates are fit to a common vertex, which must
have a fit probability greater than 0.5\%.
A mass constrained vertex fit is applied to the oppositely charged muon pairs,
in which the muon pair's mass is constrained to the known \PJGy mass 
$m_{\PJGy}$~\cite{Workman:2022ynf}, and the probability is required to be greater than 0.1\%.
The two constrained \PJGy candidates are then fit to a common 
vertex and its fit probability must exceed 0.1\%.
An event may have multiple \PJGy-pair candidates.
If both charge permutations of the same four muons have masses within 3 
standard deviations ($3\sigma$) of the known \PJGy mass, we select the pairing
that minimizes $(\Delta m_1 / \sigma_{m_1} )^2 + (\Delta m_2 / \sigma_{m_2} )^2$,
where $\Delta m$ is the difference between the reconstructed dimuon mass and
the \PJGy mass, and $\sigma_m$ is the calculated uncertainty of the reconstructed
dimuon mass (affects 0.2\% of events). All \JJ candidates are kept if they
arise from more than four muons (affects 0.2\% of events).

\begin{figure*}[!htbp]  
\centering   
\includegraphics[width=0.49\textwidth]{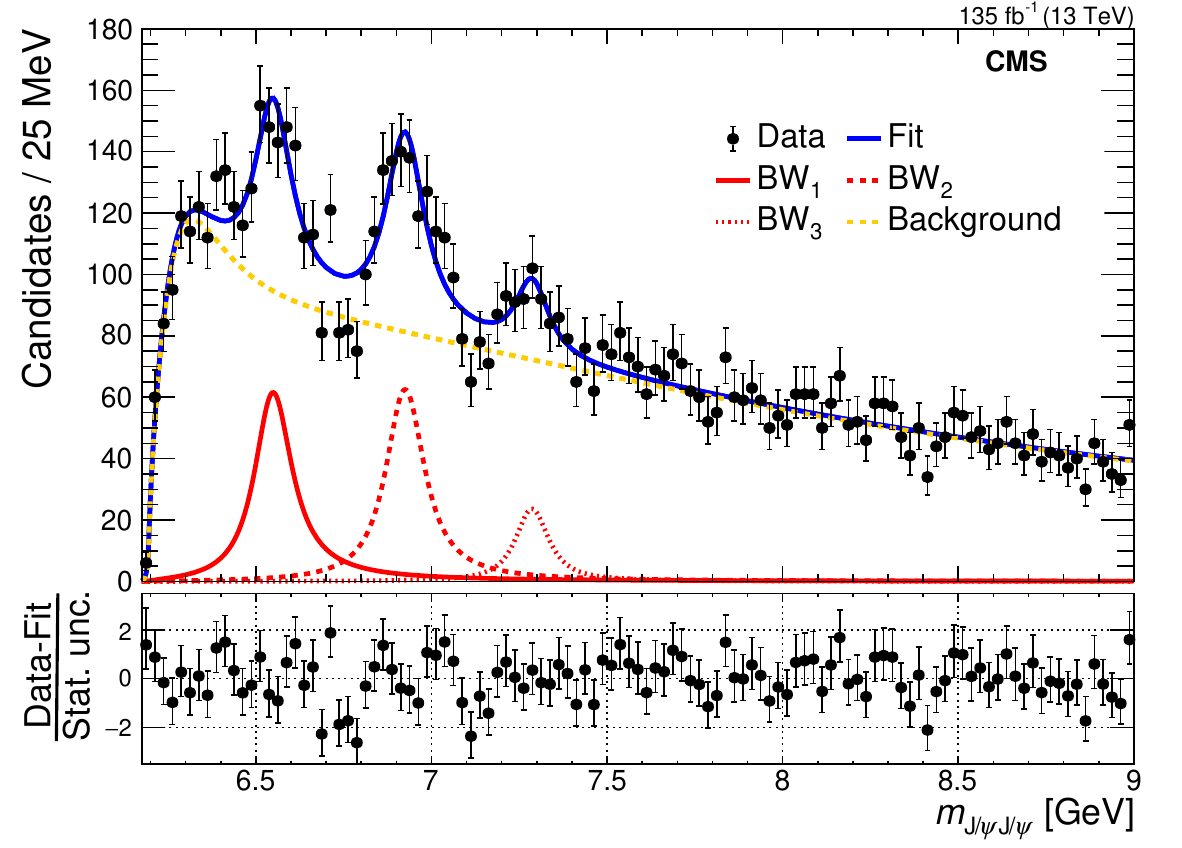} 
\includegraphics[width=0.49\textwidth]{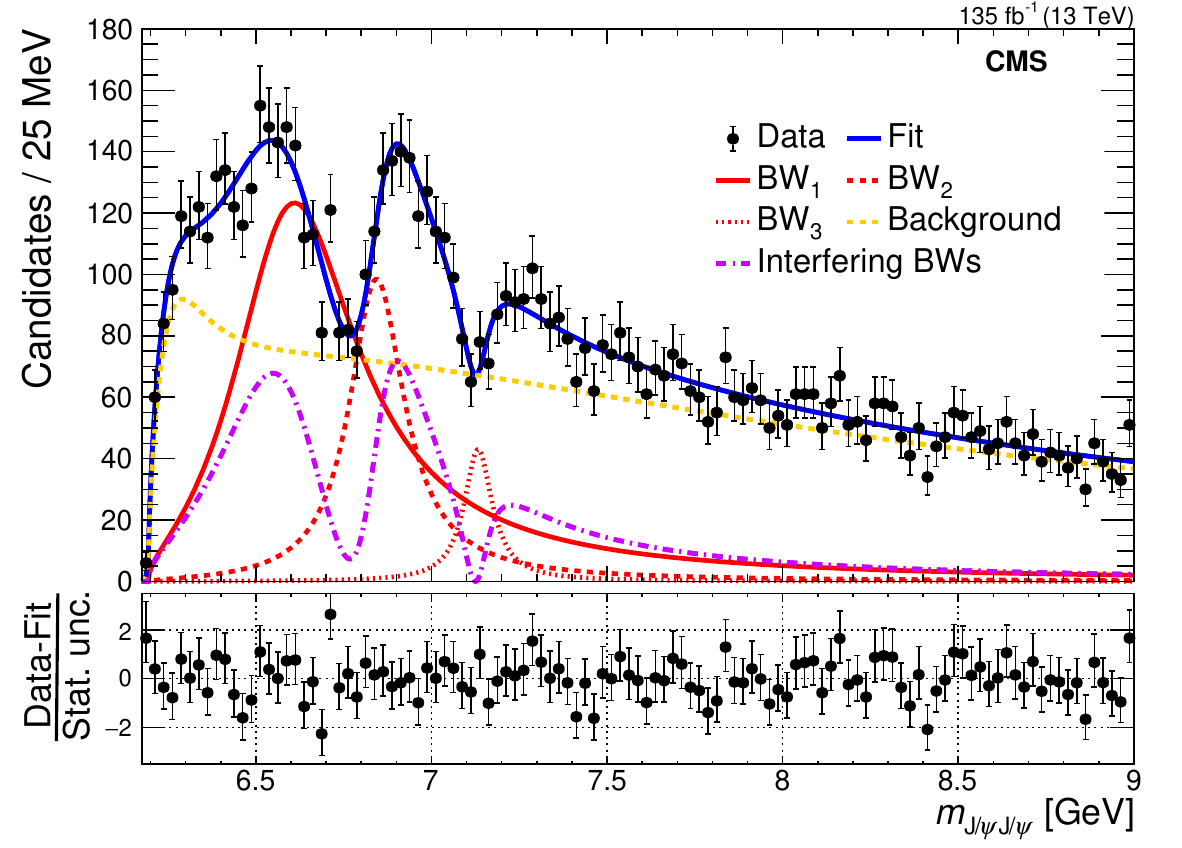} 
\caption{The \JJ invariant mass spectrum in the range up to 9\GeV, with 
fits consisting of three signal functions 
($\mathrm{BW_1}$, $\mathrm{BW_2}$, and $\mathrm{BW_3}$) and a background model (see text).
Left: the fit without interference. 
Right: the fit that includes interference,
where ``Interfering BWs'' refers to 
the total contribution of all the interfering amplitudes and their cross terms.
For clarity, only the sum of the three background components (NRSPS + DPS + $\mathrm{BW_0}$) is shown on the plots.
 The lower portion of the plots shows the pulls,
 \ie, the number of standard deviations (statistical uncertainties only)
 that the binned data differ from the fit.
}
\label{Fig:NonInterfereFit}
\end{figure*}

The \JJ invariant mass spectrum of the 8651 selected candidate pairs 
below 9\GeV is shown in Fig.~\ref{Fig:NonInterfereFit}.
To better constrain the smooth background continuum, the full analysis range 
extends up to an invariant  mass of 15\GeV
(the corresponding mass spectrum 
is shown in the Supplemental Material~\cite{suppMatBib}).

Using an unbinned extended likelihood method, 
we fit the data from the threshold $2m_{\PJGy}$ to 15\GeV. 
Signals are represented by relativistic $S$-wave Breit--Wigner (BW)
functions~\cite{Workman:2022ynf,Hippel:1972,PhysRevD.48.1225}
convolved with resolution functions, which are sums of two Gaussian functions with the same mean value. The detector mass 
resolution varies from about
10\MeV at a mass of 6500\MeV to
18\MeV at 7300\MeV. 
The BW functions are not modified by acceptance or efficiency corrections
because these 
vary only slowly, by less than 1.3\% over the natural width of the \Xsn. 

The NRSPS and DPS backgrounds are parametrizations 
of distributions of the corresponding simulated events.
The NRSPS component includes
an exponential function where one
of its parameters is a free parameter in the fit.
The effect of feed-down
from heavier mass states such as the \PGyP{2S} 
is accounted for as a systematic uncertainty.
Combinatorial backgrounds arise when one or more muon candidates are not from 
a \PJGy meson decay and from hadrons that are misidentified as a muon or 
misreconstructed. The shapes of combinatorial backgrounds are well modeled by 
the NRSPS + DPS parametrizations, and the residual effects are accounted 
for in the systematic uncertainties.

Aside from the NRSPS and DPS background components, a component to 
model a threshold enhancement is included, 
as was done in Ref.~\cite{LHCb:2020bwg}.
This excess may be due to a resonance, but other
processes could be responsible:
coupled-channel interactions~\cite{Dong:2020nwy},
triangle singularities~\cite{Guo:2019twa},
Pomeron exchange~\cite{Gong:2020bmg},
or simply an inadequate NRSPS model. 
This region could also include feed-down from  higher mass tetraquarks that
are only partially reconstructed.
Because of these uncertainties, we regard this enhancement as an additional background
component and model it with a BW function ($\mathrm{BW_0}$) with free mass and width,
which provides a good \textit{ad hoc} description of this feature.

We begin with a background-only fit
(NRSPS + DPS + $\mathrm{BW_0}$) 
and then add signal contributions one at a time to the fit, 
with the yields of each signal and background component being free parameters of the fit.
In this first series of fits, the additional components are added without interference.
A signal BW is kept as long as its 
local significance exceeds $3\sigma$, with the local significance calculated 
from the difference in log-likelihood between including and not including 
the component and only accounting for statistical uncertainties.
Three resonance structures, labeled 
$\mathrm{BW_1}$, $\mathrm{BW_2}$, 
and $\mathrm{BW_3}$ in order of increasing mass, 
are found to be statistically significant.
The local significance of these peaks, calculated from the log-likelihood 
differences between the full fit and the fit with the BW function of interest
removed, are $6.5\sigma$, $9.4\sigma$, and $4.1\sigma$, respectively.
Figure~\ref{Fig:NonInterfereFit} (left) shows this fit, 
with numerical results in Table~\ref{Tab:Fits}.
The BW function widths are all much larger than the detector mass resolution. 
Over the full mass range of 6-15\GeV, the proportions of the background 
components to the total are about 58\%, 25\%, and 9\% for NRSPS, DPS, and $\mathrm{BW_0}$, respectively.

\begin{table*}[htb]
\centering
\topcaption{Summary of the fit results for the \JJ invariant mass distribution.
 The mass $m$ and natural width $\Gamma$ 
  for both the no-interference model and the interference model,
  and the signal yields $N$ for the no-interference model, 
 are given for the three signal structures.
 The dual uncertainties are the statistical followed by the systematic components,
 and single uncertainties are only statistical.
}
\renewcommand{\arraystretch}{1.3}
\begin{scotch}{ccccc}
\hline
\hline
                &                       & $\mathrm{BW_1}$               & $\mathrm{BW_2}$           & $\mathrm{BW_3}$           \\ 
\hline
No interference & $m$ ({\MeVns})       & $6552\pm10\pm12$              & $6927\pm9\pm4$            & $7287^{+20}_{-18}\pm5$    \\   
                & $\Gamma$ ({\MeVns})   & $124^{+32}_{-26}\pm33$        & $122^{+24}_{-21}\pm18$    & $95^{+59}_{-40}\pm19$     \\ 
                & $N$                   & $470^{+120}_{-110}$           & $492^{+78}_{-73}$         & $156^{+64}_{-51}$         \\[\cmsTabSkip]
Interference    & $m$ ({\MeVns})        & $6638^{+43+16}_{-38-31}$      & $6847^{+44+48}_{-28-20}$  & $7134^{+48+41}_{-25-15}$  \\    
                & $\Gamma$ ({\MeVns})   & $440^{+230+110}_{-200-240}$   & $191^{+66+25}_{-49-17}$   & $97^{+40+29}_{-29-26}$    \\
                \hline
                \hline                
\end{scotch}
\label{Tab:Fits}
\end{table*}

We quantify the goodness of fit by the probability   
of the fit's $\chi^2$ (over the full mass range with the bin size of Fig.~\ref{Fig:NonInterfereFit}) 
and the number of degrees of freedom.
Because the signal region is only a small fraction of the fit range, 
the impact of deviations from the model in this region is diluted.
As an alternative,
we also compute a probability using 
only the bins below 7.8\GeV, which we refer to as the signal-region $\chi^2$ probability.

The full $\chi^2$ fit probability is 98\%. 
However, the dips in the data around 6750 and 7150\MeV are poorly described:
 the signal-region $\chi^2$ probability is 9\%.
Agreement may be improved by introducing interference,
as was done by LHCb~\cite{LHCb:2020bwg}.
 
Our construction of interference models is based on the three structures found in the no-interference fit.
We consider interference between pairs of (``two-way''), or all three (``three-way'') components.
The three-way interference is implemented with a term proportional to 
$\abs{r_1\exp(i\phi_1)\mathrm{BW_1} + \mathrm{BW_2} + r_3\exp(i\phi_3)\mathrm{BW_3}}^2$, where $r_{1,3}$ and 
$\phi_{1,3}$ are the relative 
magnitudes and phases 
of $\mathrm{BW_{1,3}}$ with respect to 
$\mathrm{BW_2}$~\cite{Workman:2022ynf}. 
We report the results from the three-way interference model,
which shows a signal-region 
$\chi^2$ probability of 65\%, whereas all two-way models have probabilities below 30\%.
The three-way interference fit is shown in Fig.~\ref{Fig:NonInterfereFit} (right), and 
the fit parameters are listed in Table~\ref{Tab:Fits}.

In this fit, the local significances of all three structures increase with respect to the model without interference.
For the structure with the lowest significance ($\mathrm{BW_3}$), a value of $4.7\sigma$ is obtained.
The global significance of   $\mathrm{BW_3}$ is estimated by 
generating  MC pseudoexperiments and determining
the probability that a statistical fluctuation yields a  ``signal'' 
whose local significance equals or exceeds that of $\mathrm{BW_3}$
in the windows   7.05--7.8\GeV for mass and 50--260\MeV for width.
The $\mathrm{BW_3}$ phase is unconstrained. 
The resulting global significance for $\mathrm{BW_3}$
is  $3.4\sigma$.

The interference between different resonances is motivated by the idea that the states could have the same quantum 
numbers and be coherently produced. One can also consider interference between the resonances and the NRSPS background. 
We consider this scenario less probable
as the background is likely a mixture of $J^{PC}$ states. We have investigated 
two models with this sort of interference. The first model is identical to Model II from Ref.~\cite{LHCb:2020bwg} 
with a signal BW function for the \Xsn, an auxiliary BW function around 6700\MeV, the NRSPS and DPS background, and interference 
between the NRSPS background and auxiliary BW function. This did not provide a good fit to our data 
as shown in the Supplemental Material~\cite{suppMatBib}.
The second model starts from the three-way interference model and adds interference between the three 
signal BW functions and the NRSPS background. This model did not significantly improve the fit quality.

While interference between resonances is one possible mechanism, other models may also be able to 
reproduce the dips. For example, inspired by predictions of dense spectroscopy of tetracharm 
states~\cite{Wang:2020wrp,Chen:2020xwe,Bedolla:2019zwg,Jin:2020jfc,liu:2020eha,Liu:2021rtn,Zhu:2020xni}, 
representing individual peaks by multiple overlapping narrow resonances
could provide good fits by enabling the BW functions to fall more deeply into the dips.

Systematic uncertainties for masses and widths are determined by varying aspects of,
and inputs to, the fits.
For a given source, the largest deviation from a parameter's nominal value is taken as its systematic uncertainty.
The sources of systematic uncertainty that have been considered are as follows:
 different BW function shapes ($P$ and $D$ wave, and alternative values for the ``interaction radius'' in the Blatt--Weisskopf barrier factor~\cite{Hippel:1972, PhysRevD.48.1225});
alternative DPS parametrizations, such as those obtained by mixing data events
 (artificial \JJ mass spectrum formed by two $\PJGy$s from different events);
 NRSPS parametrizations from different simulations
(\CASCADE and \HELACONIA) or floating otherwise fixed parameters individually;
 momentum scale (based on the shift in the unconstrained fitted \PJGy meson mass 
 from the world average value~\cite{Workman:2022ynf});
 detector mass resolution (slightly different resolution models based on the \PYTHIA or \JHUGEN generators);
combinatorial background shape 
(the result of altering the parameterization of the combinatorial background);
efficiency corrections 
(the difference between not applying and applying efficiency corrections
based on the \PYTHIA or \JHUGEN generator);
and the effects of including various feed-down components from hypothetical heavier tetraquarks~\cite{Wang:2020tpt}. 
The effects of feed-down components in the interference model can produce 
 asymmetric uncertainties, so an asymmetric uncertainty 
is assigned for this source. For other sources, symmetric uncertainties 
are assigned.
The principal systematic effects are summarized in Table~\ref{Tab:Sys}.
The total uncertainties are their sum in quadrature
and also appear in Table~\ref{Tab:Fits}.
Including feed-down in the fit affects signal structures, especially 
for the $\mathrm{BW_1}$, 
because of their overlap in the \JJ mass distribution.
Compared to the no-interference model, the greater complexity and increased parameter correlations of the interference 
model result in larger 
systematic uncertainties.

\begin{table*}[!htbp]\centering
\topcaption{Dominant contributions to the systematic uncertainties in masses and widths, in \MeV, 
for the two fits. The ``Total uncertainty'' is the quadratic sum of all individual components, 
including the unlisted nondominant contributions.}
 \begin{scotch}{llrrrrrr}
     \hline
     \hline 
Fit             & Dominant sources      & $M_{\mathrm{BW_1}}$   & $M_{\mathrm{BW_2}}$   & $M_{\mathrm{BW_3}}$   & $\Gamma_{\mathrm{BW_1}}$  & $\Gamma_{\mathrm{BW_2}}$  & $\Gamma_{\mathrm{BW_3}}$  \\[\cmsTabSkip]
\hline
No-interference & Signal shape          & 3                     & 3                     & 3                     & 10                        & 5                         & 5                         \\
                & NRSPS shape           & 3                     & 1                     & 1                     & 18                        & 15                        & 17                        \\
                & Feed-down             & 11                    & 1                     & 1                     & 25                        & 8                         & 6                         \\[\cmsTabSkip]
                & Total uncertainty     & 12                    & 4                     & 5                     & 33                        & 18                        & 19                        \\[\cmsTabSkip] 
Interference    & Signal shape          & 7                     & 12                    & 7                     & 56                        & 8                         & 7                         \\ 
                & DPS shape             & 1                     & 3                     & 2                     & 18                        & 6                         & 2                         \\
                & NRSPS shape           & 9                     & 14                    & 13                    & 85                        & 9                         & 20                        \\
                & Mass resolution       & 8                     & 4                     & 1                     & 24                        & 7                         & 13                        \\
                & Combinatorial background    & 7                     & 2                     & $<$1                  & 5                         & 3                         & 2                         \\
                & Feed-down             & \Large $^{+0}_{-27}$  & \Large $^{+44}_{-0}$  & \Large $^{+38}_{-0}$  & \Large $^{+0}_{-210}$     & \Large $^{+19}_{-0}$      & \Large $^{+12}_{-0}$      \\[\cmsTabSkip]
                & Total uncertainty     & \Large $^{+16}_{-31}$ & \Large $^{+48}_{-20}$ & \Large $^{+41}_{-15}$ & \Large $^{+110}_{-240}$   & \Large $^{+25}_{-17}$     & \Large $^{+29}_{-26}$    \\
                \hline
                \hline
               \end{scotch}
  \label{Tab:Sys}
\end{table*}
 
The impact of systematic uncertainties on the local
significances of the structures
was checked in the interference fit model 
by using discrete sets of individual alternative hypotheses and recomputing
the significances. The systematic uncertainties 
introduce no appreciable change.

The $\mathrm{BW_2}$ parameters are within 2$\sigma$, when comparing the 
analogous no-interference or interference models, 
of the \Xsn values reported by LHCb. 
For their model~I, consisting of a BW function for the \Xsn signal plus two auxiliary 
threshold BW functions, they reported an \Xsn mass 
of $6905\pm 11 \pm 7\MeV$ with a natural width 
of $80\pm 19 \pm 33\MeV$, and $6886\pm 11 \pm 11\MeV$ and 
$168\pm 33 \pm 69\MeV$ for model~II~\cite{LHCb:2020bwg}.
However, because the two experiments use different fit models,
quantitative comparisons between them are not straightforward, and 
thus we also fit our data with the two LHCb models. 
Neither LHCb model provides a good description of the CMS data; 
nevertheless, our \Xsn parameters are again compatible.
More information about these comparisons, including where some small discrepancies are present,
is summarized in the Supplemental Material~\cite{suppMatBib}.

Our measured masses appear compatible with
recent calculations of the $\PQc\PQc\PAQc\PAQc$
spectrum~\cite{Tiwari:2021tmz,Zhu:2020xni},
which would indicate that these three structures may be a family of radial excitations
of the same $J^{PC}$.
This is the case for both no-interference and interference masses,
albeit for different theoretical models.

In summary, the \JJ invariant mass spectrum has been presented. 
The data were collected with the CMS detector from $\Pp\Pp$ collisions at $\sqrt{s} = 13\TeV$ 
and correspond to an integrated luminosity of 135\fbinv. 
Three structures are found in the \JJ invariant mass spectrum.
The spectrum is better described by a model with interference among three resonances.
Two new structures, tentatively named \Xss
and   \Xst, 
are found with masses of
     $6638^{+43}_{-38} \stat ^{+16}_{-31} \syst$ and  
     $7134^{+48}_{-25} \stat ^{+41}_{-15} \syst\MeV$;
and the \Xsn structure 
observed 
by LHCb is confirmed with a mass 
of $6847^{+44}_{-28} \stat ^{+48}_{-20} \syst\MeV$.
The local statistical 
significances of these peaks
are, for increasing mass,
7.9, 9.8, and 4.7 standard deviations.
Numeric results are provided in the HEPData record~\cite{hepdata}.

\textit{Note added.}\textemdash Recently, the ATLAS Collaboration released a Letter~\cite{ATLAS:2023bft} 
confirming the \Xsn structure in the \JJ spectrum 
and also reported near-threshold excesses, including 
a possible feature in the $\PGyP{2S}\PJGy$ channel.

\begin{acknowledgments}
We congratulate our colleagues in the CERN accelerator departments for the excellent performance of the LHC and thank the technical and administrative staffs at CERN and at other CMS institutes for their contributions to the success of the CMS effort. In addition, we gratefully acknowledge the computing centers and personnel of the Worldwide LHC Computing Grid and other centers for delivering so effectively the computing infrastructure essential to our analyses. Finally, we acknowledge the enduring support for the construction and operation of the LHC, the CMS detector, and the supporting computing infrastructure provided by the following funding agencies: BMBWF and FWF (Austria); FNRS and FWO (Belgium); CNPq, CAPES, FAPERJ, FAPERGS, and FAPESP (Brazil); MES and BNSF (Bulgaria); CERN; CAS, MoST, NSFC, Nanjing Normal University research start-up funding project,  Tsinghua University Initiative Scientific Research and Dushi Programs, and Natural Science Foundation of China under Grants No. 11975010, No. 12075123, and No. 12061141002 (China); MINCIENCIAS (Colombia); MSES and CSF (Croatia); RIF (Cyprus); SENESCYT (Ecuador); MoER, ERC PUT and ERDF (Estonia); Academy of Finland, MEC, and HIP (Finland); CEA and CNRS/IN2P3 (France); BMBF, DFG, and HGF (Germany); GSRI (Greece); NKFIH (Hungary); DAE and DST (India); IPM (Iran); SFI (Ireland); INFN (Italy); MSIP and NRF (Republic of Korea); MES (Latvia); LAS (Lithuania); MOE and UM (Malaysia); BUAP, CINVESTAV, CONACYT, LNS, SEP, and UASLP-FAI (Mexico); MOS (Montenegro); MBIE (New Zealand); PAEC (Pakistan); MES and NSC (Poland); FCT (Portugal); MESTD (Serbia); MCIN/AEI and PCTI (Spain); MOSTR (Sri Lanka); Swiss Funding Agencies (Switzerland); MST (Taipei); MHESI and NSTDA (Thailand); TUBITAK and TENMAK (Turkey); NASU (Ukraine); STFC (United Kingdom); DOE and NSF (U.S.).
\end{acknowledgments}
\bibliography{auto_generated} 

\ifthenelse{\boolean{cms@external}}{}{
\clearpage
\appendix
\numberwithin{table}{section}
\numberwithin{figure}{section}
\section{Fit projections in full range\label{app:suppMat}}
\input{supplemental_material}

}
\cleardoublepage \section{The CMS Collaboration \label{app:collab}}\begin{sloppypar}\hyphenpenalty=5000\widowpenalty=500\clubpenalty=5000\input{BPH-21-003-public-authorlist.tex}\end{sloppypar}
\end{document}

%% file: supplemental_material.tex
\begin{figure*}[!htbp] 
\centering    
\includegraphics[width=0.98\textwidth]{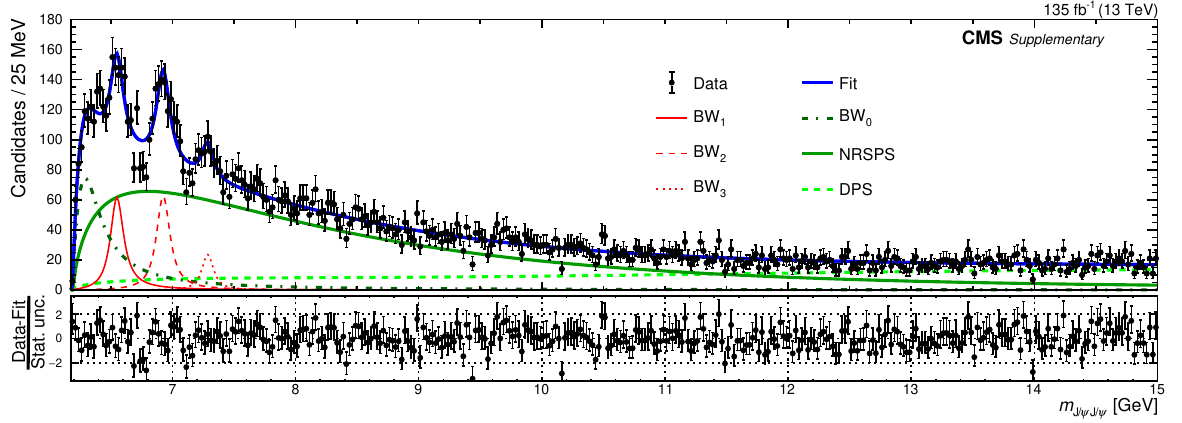}
\includegraphics[width=0.98\textwidth]{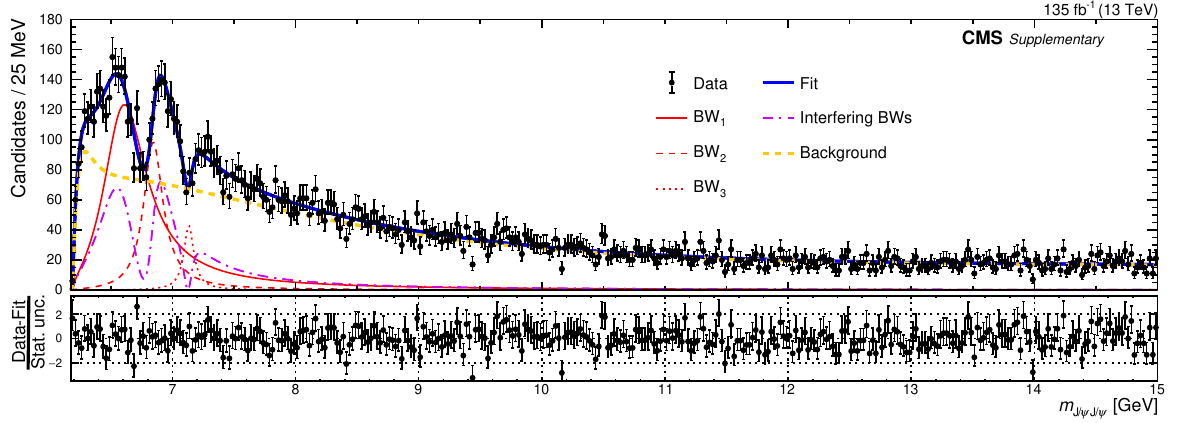}
\caption{The \JJ invariant mass spectrum with the no-interference fit (upper) and
the interference fit (lower) in the full fit range 
(see text for model details).
The ``Interference BWs'' curve is the total contribution of
all the interference amplitudes and their cross terms.
The lower portion of the plots shows the pulls, 
 \ie, the number of standard deviations (statistical uncertainties only) 
 that the binned data differ from the fit.
}
\label{Fig:full_range}
\end{figure*}

The \JJ invariant mass spectrum
for the full range of the fit 
is shown in Fig.~\ref{Fig:full_range} for the 
no-interference and the interference models.

\section{Fitting with the LHCb models\label{app:suppMatB}}

In order to make a direct comparison with the results from the LHCb
Collaboration~\cite{LHCb:2020bwg} 
we also used their two principal models (I and II) to fit the CMS data.
The details are provided here.

Model I consists of two auxiliary relativistic Breit--Wigner (BW) functions
below the \Xsn BW.
The parameter values for these two BWs are not reported in Ref.~\cite{LHCb:2020bwg}, 
but the second auxiliary BW is in the same mass region as the CMS ``$\mathrm{BW_1}$''
and so we extend this labeling to the LHCb models.

\begin{figure*}[!htbp] 
\centering   
\includegraphics[width=0.98\textwidth]{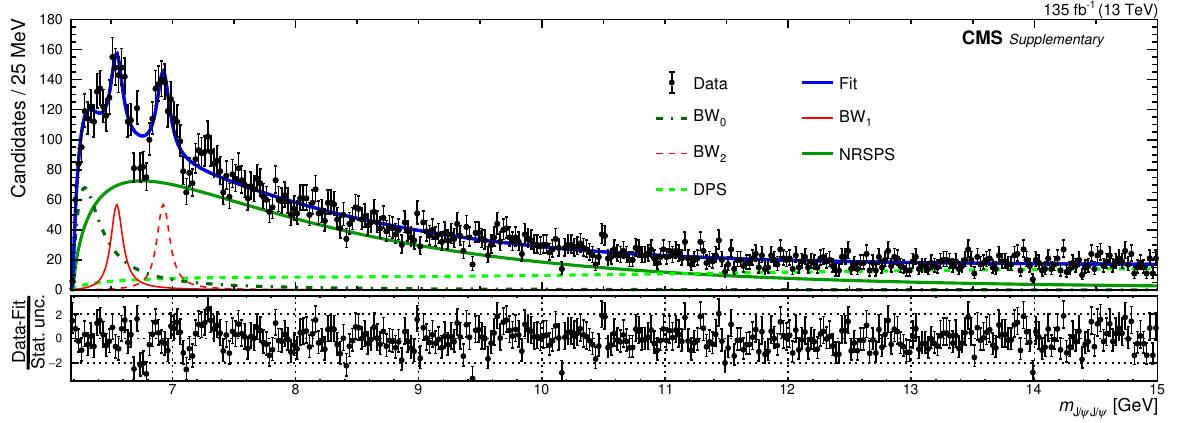}
\includegraphics[width=0.98\textwidth]{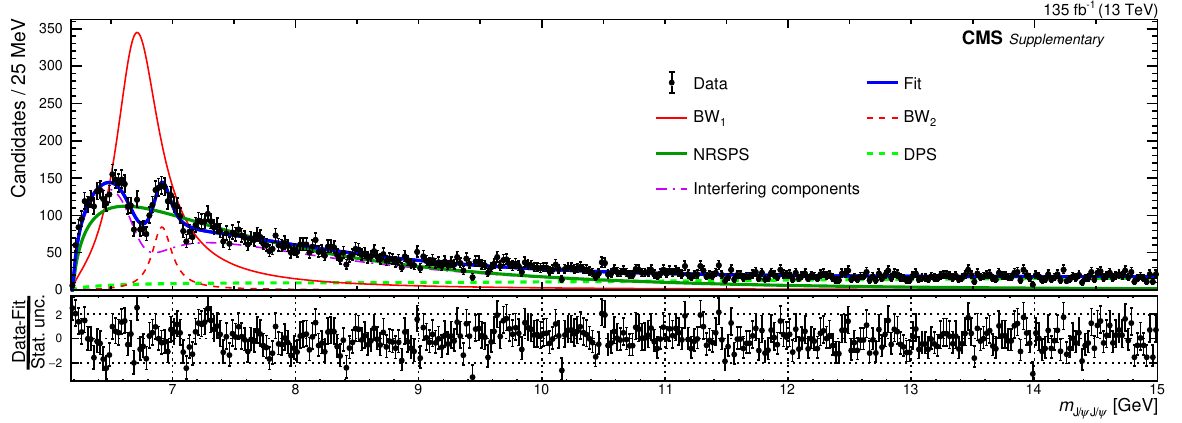}
\caption{The \JJ invariant mass spectrum with fits using LHCb models~\cite{LHCb:2020bwg}: 
Model I (no-interference) (upper) and Model II (interference) (lower).
Model details are given in the text.
The ``Interfering components'' is the total contribution of $\mathrm{BW_{1}}$, NRSPS, and their interference.
 The lower portion of the plots shows the pulls
 between the binned data and the fit.}
\label{Fig:LHCbLike}
\end{figure*}

The Model~I fit is shown in Fig.~\ref{Fig:LHCbLike} (upper),
with numerical results listed in Table~\ref{Tab:LHCbFits}.
The fit quality is poor with a signal-region $\chi^2$ probability of 0.9\%.
The no-interference fit using the CMS model with three structures---and roughly the same number 
of \JJ candidates after the final selection as LHCb---has a signal-region 
$\chi^2$ probability of 9\%.

 \begin{table*}[!htbp]
\centering
\topcaption{Summary of results for the two LHCb fit models from LHCb and CMS: 
  the mass $m$ and natural width $\Gamma$, in \MeV,  
  are given for the principal structures
  (the Breit--Wigner function at the threshold is not included in Model~I),
  and LHCb values not reported in Ref.~\cite{LHCb:2020bwg} are marked as ``\NA''. 
  Single uncertainties are statistical only.
   }

 \begin{scotch}{cccccc}
 Exp. &  Fit    &     $M_{\mathrm{BW_{1}}}$  & $\Gamma_{\mathrm{BW_{1}}}$  &     $M_{\Xsn}$   & $\Gamma_{\Xsn}$    \\ 
\hline
 LHCb~\cite{LHCb:2020bwg}   & Model I  &  \NA & \NA & $6905 \pm 11 \pm 7$ &   $80 \pm 19 \pm 33$ \\
 CMS                        & Model I  & $6550 \pm 10$ &  $112 \pm 27$  &  $6927 \pm 10$  & $117 \pm 24$ \\ 
 LHCb~\cite{LHCb:2020bwg}   & Model II & $6741 \pm 6$  & $ 288 \pm 16$ & $6886 \pm 11  \pm 11$ &   $168 \pm 33 \pm 69$ \\
 CMS                        & Model II & $6736 \pm 38$ &  $ 439 \pm 65$  &  $6918 \pm 10$  & $187 \pm 40$
\end{scotch}
\label{Tab:LHCbFits}
\end{table*}

Model~II has, aside from the signal BW, a single auxiliary BW (here labeled as `$\mathrm{BW_1}$'), 
which interferes with the NRSPS background.
This interference was introduced to better model the dip around 6750\MeV.
Our fit results for Model~II are shown in Fig.~\ref{Fig:LHCbLike} (lower)
and Table~\ref{Tab:LHCbFits}.
The 6750\MeV dip is now well described: the pulls in Fig.~\ref{Fig:LHCbLike} (lower)
are well behaved in the vicinity of the \Xsn peak.
However, systematic deviations are evident elsewhere in the pulls,
and the signal-region $\chi^2$ probability of 0.8\% for this fit
is actually lower than for Model~I.

%% file: BPH-21-003-public-authorlist.tex
\cmsinstitute{Yerevan Physics Institute, Yerevan, Armenia}
{\tolerance=6000
A.~Hayrapetyan, A.~Tumasyan\cmsAuthorMark{1}\cmsorcid{0009-0000-0684-6742}
\par}
\cmsinstitute{Institut f\"{u}r Hochenergiephysik, Vienna, Austria}
{\tolerance=6000
W.~Adam\cmsorcid{0000-0001-9099-4341}, J.W.~Andrejkovic, T.~Bergauer\cmsorcid{0000-0002-5786-0293}, S.~Chatterjee\cmsorcid{0000-0003-2660-0349}, K.~Damanakis\cmsorcid{0000-0001-5389-2872}, M.~Dragicevic\cmsorcid{0000-0003-1967-6783}, A.~Escalante~Del~Valle\cmsorcid{0000-0002-9702-6359}, P.S.~Hussain\cmsorcid{0000-0002-4825-5278}, M.~Jeitler\cmsAuthorMark{2}\cmsorcid{0000-0002-5141-9560}, N.~Krammer\cmsorcid{0000-0002-0548-0985}, D.~Liko\cmsorcid{0000-0002-3380-473X}, I.~Mikulec\cmsorcid{0000-0003-0385-2746}, J.~Schieck\cmsAuthorMark{2}\cmsorcid{0000-0002-1058-8093}, R.~Sch\"{o}fbeck\cmsorcid{0000-0002-2332-8784}, D.~Schwarz\cmsorcid{0000-0002-3821-7331}, M.~Sonawane\cmsorcid{0000-0003-0510-7010}, S.~Templ\cmsorcid{0000-0003-3137-5692}, W.~Waltenberger\cmsorcid{0000-0002-6215-7228}, C.-E.~Wulz\cmsAuthorMark{2}\cmsorcid{0000-0001-9226-5812}
% ...
\par}
\vskip\cmsinstskip
\dag:~Deceased\\
$^{1}$Also at Yerevan State University, Yerevan, Armenia\\
$^{2}$Also at TU Wien, Vienna, Austria\\
% ...
